\documentclass[journal,hideappendix]{vgtc}        

\PassOptionsToPackage{svgnames,dvipsnames}{xcolor}

\onlineid{1199}

\vgtccategory{Research}

\title{Mystique: Deconstructing SVG Charts for Layout Reuse}

\author{\authororcid{Chen Chen}{0000-0003-3171-0657}, \authororcid{Bongshin Lee}{0000-0002-4217-627X}, \authororcid{Yunhai Wang}{0000-0003-0059-6580},  Yunjeong Chang, and \authororcid{Zhicheng Liu}{0000-0002-1015-2759}}

\authorfooter{
  \item
  	Chen Chen, Zhicheng Liu, and Yunjeong Chang are with University of Maryland, College Park, Maryland, United States.
  	E-mail: \{cchen24, leozcliu\}@umd.edu, ychangs2@terpmail.umd.edu.

  \item Bongshin Lee is with Microsoft Research, Redmond, Washington, United States.
  	E-mail: bongshin@microsoft.com.

     \item
  	Yunhai Wang is with Shandong University, Qingdao, Shandong, China.
  	E-mail: cloudseawang@gmail.com.
}

\abstract{To facilitate the reuse of existing charts, previous research has examined how to obtain a semantic understanding of a chart by deconstructing its visual representation into reusable components, such as encodings. However, existing deconstruction approaches primarily focus on chart styles, handling only basic layouts. In this paper, we investigate how to deconstruct chart layouts, focusing on rectangle-based ones, as they cover not only 17 chart types but also advanced layouts (e.g., small multiples, nested layouts). We develop an interactive tool, called Mystique, adopting a mixed-initiative approach to extract the axes and legend, and deconstruct a chart's layout into four semantic components: mark groups, spatial relationships, data encodings, and graphical constraints. Mystique employs a wizard interface that guides chart authors through a series of steps to specify how the deconstructed components map to their own data. On 150 rectangle-based SVG charts, Mystique achieves above 85\% accuracy for axis and legend extraction and 96\% accuracy for layout deconstruction. In a chart reproduction study, participants could easily reuse existing charts on new datasets. We discuss the current limitations of Mystique and future research directions.}


\teaser{
  \centering
  \includegraphics[width=\linewidth]{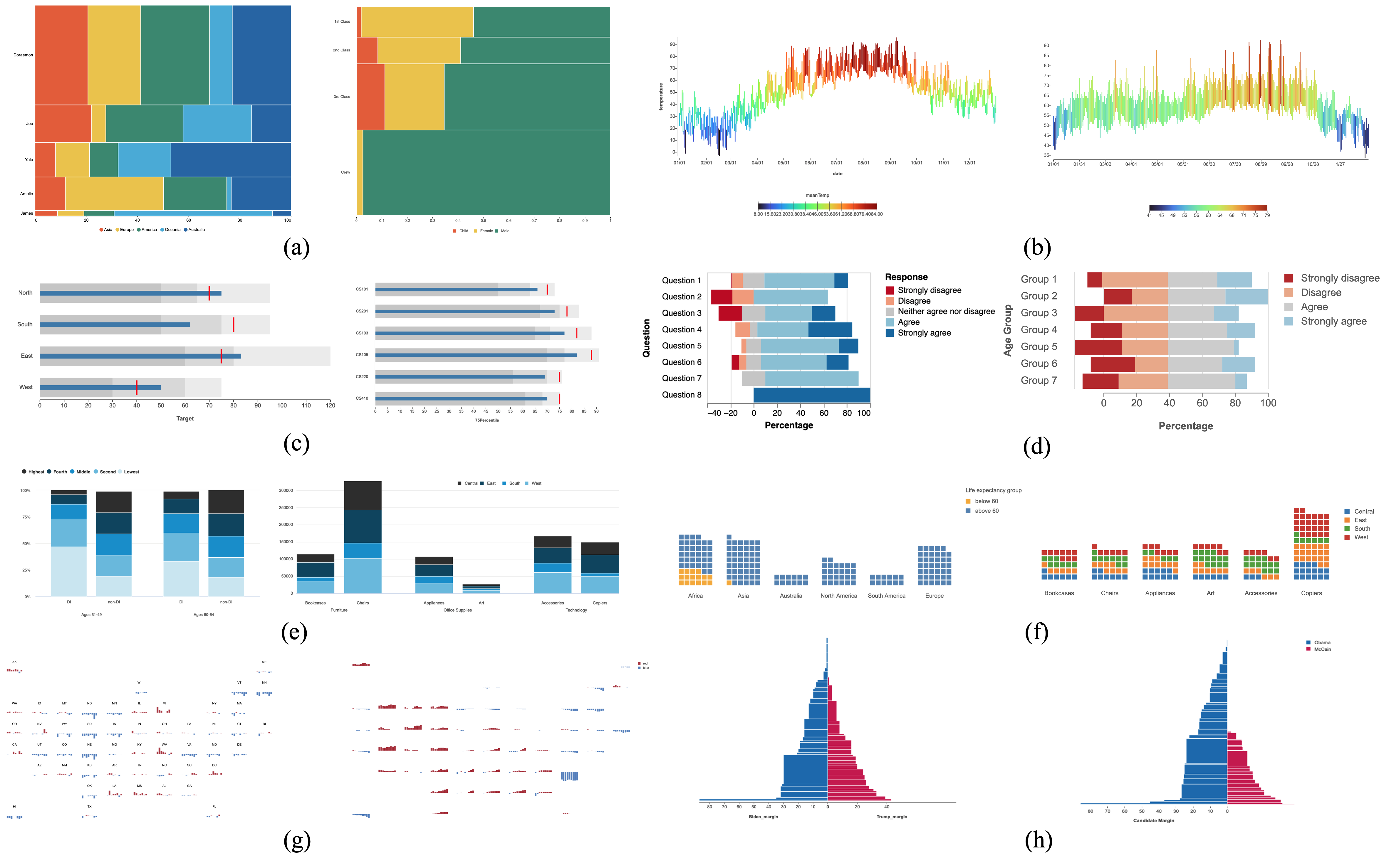}
  \caption{Each pair shows an existing chart (left) and a new chart created using Mystique (right). The existing charts are produced using a variety of tools, such as D3, Vega-lite, Mascot, PlotDB, Highcharts, and Data Illustrator.}
  \label{fig:teaser}
}

\usepackage[dvipsnames]{xcolor}
\definecolor{NavyBlue}{HTML}{000080}
\graphicspath{{figs/}{figures/}{pictures/}{images/}{./}} 
\usepackage{multirow}
\renewcommand{\arraystretch}{1.5}
\usepackage{makecell}
\usepackage{colortbl}
\usepackage{booktabs}

\usepackage{wrapfig}
\usepackage{xspace}

\usepackage[skip=5pt]{caption}
\usepackage{subcaption}
\usepackage{pifont}

\usepackage{url}
\usepackage{ textcomp }
\usepackage{svg}
\usepackage{amsmath}

\newcommand{\bpstart}[1]{\vspace{1mm} \noindent{\textbf{#1.}}}
\newcommand{\bstart}[1]{\vspace{1mm} \noindent{\textbf{#1}}}
\newcommand{\eg}{{e.g.,}\xspace}

\newcommand{\etal}{{et al.}\xspace}
\newcommand{\revise}[1]{\textcolor{black}{#1}}
\newcommand{\markup}[1]{\textcolor{black}{#1}}

\usepackage{tabu}                      
\usepackage{booktabs}                  
\usepackage{lipsum}                    
\usepackage{mwe}                       
\usepackage{mathptmx}                  
\usepackage{stfloats} 

\begin{document}

\definecolor{tableheader}{HTML}{EFEFEF}
\definecolor{tablegrayline}{HTML}{d0d0d0}



\maketitle

\section{Introduction}\label{sec:intro}

Data visualization creators often look for and incorporate existing visualizations in their work \cite{bako2022understanding,bigelow_reflections_2014}, because these visualizations serve as concrete examples that embody design ideas in terms of encodings, visual styles, and layouts~\cite{bako2022understanding}. Even though numerous visualizations are available on the online galleries of visualization languages (\eg~D3 \cite{bostock_d3_2011}, Observable Plot~\cite{observableplot}, Vega-lite \cite{satyanarayan_vega-lite_2016}) and authoring tools (\eg~Tableau Public \cite{tableau}, Data Illustrator \cite{liu_data_2018}, Charticulator \cite{ren_charticulator_2018}), it remains a major challenge to re-purpose these visualizations with users' own datasets in the chart authoring workflow \cite{bako2022understanding, Battle2021D3}.  Users have difficulties specifying the mappings between their data and different components in a visualization using programming constructs provided in a language \cite{bako2022understanding, Battle2021D3}, or they have to understand the underlying framework or grammar in an authoring tool and create a visualization starting from an empty canvas \cite{satyanarayan_critical_2019}. 

Researchers thus have explored how to enable the reuse of existing visualizations without requiring users to learn a new language or start from scratch. For example, D3 Deconstructor \cite{harper_converting_2017,harper_deconstructing_2014} turns basic D3 charts into reusable style templates, and Chartreuse \cite{cui_mixed-initiative_2022} supports the reuse of infographics bar charts in Microsoft PowerPoint. A central challenge in these works is to obtain a semantic understanding of a chart by deconstructing its visual representation into components (e.g., encodings) that can be reused with a new dataset. Existing approaches to deconstructing charts for reuse, however, are limited because of their primary focus on chart styles instead of layouts. D3 Deconstructor \cite{harper_converting_2017} extracts style templates only from basic charts (\eg bar charts, scatter plots), where the spatial arrangements of marks can be described using simple data bindings. Chartreuse \cite{cui_mixed-initiative_2022} focuses on glyphs with visual styles only in bar charts with simple layouts. As a result, complex visualizations such as small multiples and charts with nested layouts are not supported. In addition, existing approaches for chart deconstruction often require that charts are created using specific tools like D3.js \cite{bostock_d3_2011} or PowerPoint \cite{cui_mixed-initiative_2022}, further limiting the range of reusable visualizations.

To address these limitations, we extend existing work and investigate how to deconstruct SVG (Scalable Vector Graphics) charts for layout reuse. 
We selected SVG as the input format for two reasons: (1) it is a tool-agnostic standard image format for 2-dimensional graphics and is supported in most charting libraries and systems, allowing abundant sources of reusable charts and (2) unlike raster images where the visual marks need to be segmented and extracted, every mark in an SVG chart is specified as an individual SVG element~\cite{masson2023chartdetective}.

\revise{We started our investigation with the Beagle dataset~\cite{battle2018beagle}, which \markup{samples online SVG charts and analyzes their distributions.}
We found that line-, circle-, and pie-based charts all have only a few variants~(\cref{tbl:lineCircleRect}) and their layouts are simple in that positions of marks are usually determined by data bindings. Furthermore, they are supported in most charting tools as reusable templates. It was also observed that charts composed of other marks like area and polygon account for a small portion~($\approx$5\%).} 
Thus, the benefits of deconstructing and reusing these charts are marginal. 
In contrast, charts composed of rectangles encompass many more chart types and afford expressive and diverse designs regarding chart layouts. The layouts of many rectangle-based charts are determined by multiple factors beyond simple data binding.
For example, the positions of the bars in~\cref{fig:teaser}d are determined by the stacking of the bars and the vertical placement \& alignment of bar groups; the position of a rectangle in \cref{fig:teaser}g depends on its position within the corresponding bar group, whose position encodes data.

In such cases, it is often difficult to identify and specify the different factors determining the layout, yet few charting tools offer these charts as standard templates. To the best of our knowledge, no work has examined how to tease apart underlying factors that jointly determine these chart layouts for reuse. We thus focus on rectangle-based charts because they present interesting research challenges and offer benefits to users who want to create similar charts without having to learn a new visualization framework or language. 
Specifically, we seek to answer the following two research questions.
\textbf{RQ1}: How can we deconstruct rectangle-based SVG charts into layout components that jointly determine marks' spatial positions?
\textbf{RQ2}: How do we apply the deconstruction result from an SVG chart to a new dataset for reuse?

To answer RQ1, we introduce 
\textbf{mixed-initiative chart deconstruction} by combining automated algorithms with user input. Starting with the automatic extraction of axes and legend, we provide an interface for a user to correct any mistakes in the extraction results through simple interactions. A hierarchical clustering algorithm then decomposes the main chart content into four semantic components that determine the chart layout: \revise{\textit{grouping}, \textit{spatial relationships}}, \textit{encodings}, and \textit{graphical constraints}. Compared to previous approaches \cite{harper_converting_2017,cui_mixed-initiative_2022,chen_towards_2020}, our approach works on SVG charts with nested or bespoke structures that are created by a wide range of visualization tools, and can handle errors and uncertainties in the deconstruction pipeline.

To answer RQ2, we propose \textbf{guided chart reuse through a wizard interface}. The reuse process starts with the user importing a dataset, where its compatibility with the example chart is checked. A wizard interface then guides the user through a series of steps to specify mappings between data attributes and visual objects or channels, which leads to a new visualization for the user's dataset.
Furthermore, the generated chart is in a format compatible with an existing authoring tool (i.e., Data Illustrator~\cite{liu_data_2018}), enabling further interactive customization without the need to program.

We develop these solutions in an interactive prototype, Mystique, and evaluate the viability of our approach in two ways. We demonstrate that \revise{our method advances the state of the art in chart deconstruction with over 96\% accuracy on 150 real-world SVG charts produced by 25 different tools, covering not only 17 chart types but also advanced layouts such as small multiples and nested grouping.} 
We also conduct a chart re-production study with $12$ participants to evaluate the usability of the wizard interface. The participants were able to create new charts in a few minutes with Mystique by reusing the charts given to them.

\begin{table}[t]
\caption{Types and percentages of charts composed of lines, circles, pies, arcs, rectangles, and other marks in the Beagle dataset~\cite{battle2018beagle}.}
    \centering
    {\small
    \setlength\tabcolsep{1.5pt}
    \begin{tabular}{lp{0.725\linewidth}r}
    \toprule
     Mark   & Chart  & Percentage\\
     \midrule
      Rectangle & bar chart~(histogram), grouped bar chart, stacked bar chart, diverging bar chart (pyramid chart), Marimekko chart, heatmap, bullet chart, treemap, waffle chart, waterfall chart, range chart, gantt chart, matrix chart, cartogram, calendar chart & 32.85\%\\
      Line  &  line graph, parallel coordinates, Kagi chart &  30.51\%\\
      Pie &  pie chart, donut chart & 16.50\%\\
      Circle &  scatter plot, bubble plot, dot plot, circle packing & 14.96\%\\
      Others & geographic map, area chart, stream graph, chord chart, hexbin plot, Sankey diagram, Voronoi diagram, word cloud, sunburst chart, boxplot, network diagram, contour plot, radial plot & 5.18\%\\
     \bottomrule
    \end{tabular}
    }
    \label{tbl:lineCircleRect}
\end{table}

\section{Related Work}   
\subsection{\revise{Chart Reuse} Approaches}\label{sec:2.1}
\revise{To create new charts, previous studies on visualization designers' practices~\cite{bigelow_reflections_2014,walny_data_2019} suggest that it is more natural to change existing graphics than to start from scratch.}
Templates are generally recognized as a user-friendly way to create charts, especially for beginners. In traditional template-based systems, templates are created by system developers and they usually suffer from limited expressivity and quantity. 
Previous research thus has investigated how to turn existing visualizations into reusable templates without involving developers. For example, D3 Deconstructor \cite{harper_deconstructing_2014,harper_converting_2017} works on basic charts created using D3.js; iVoLVER \cite{mendez_ivolver_2016} extracts data from charts and updates them with new data; 
Ivy \cite{mcnutt_integrated_2021} supports turning JSON-based declarative specifications into parameterized templates; 
Chen~\etal~\cite{chen_towards_2020} use deep learning to extract timelines from infographics as templates; and Chartreuse \cite{cui_mixed-initiative_2022} supports reusing infographics bar chart templates. 

Overall, D3 Deconstructor and Chartreuse are the closest work to Mystique. D3 Deconstructor only takes charts created using D3 \cite{bostock_d3_2011}, which have the source data embedded, and Chartreuse primarily works on Microsoft PowerPoint graphics assets. In contrast, Mystique works on visualizations in the general SVG format, does not require access to underlying data, \revise{and supports more advanced layouts}. 



\subsection{Chart Understanding and Deconstruction}
Making a visualization example reusable requires understanding and deconstructing visualizations.  
Various automated or semi-automated methods have been proposed to detect marks \cite{ying_glyphcreator_2022,chen_towards_2020} as well as axes and legends \cite{shukla_recognition_2008, choudhury_scalable_2016}, classify chart types \cite{savva_revision_2011,shukla_recognition_2008}, and extract data \cite{jung_chartsense_2017,harper_converting_2017,harper_deconstructing_2014,masson2023chartdetective} and visual encodings \cite{poco_reverse-engineering_2017,poco_extracting_2018,harper_deconstructing_2014,harper_converting_2017,cui_mixed-initiative_2022}. Due to the vast space of visualization examples, these methods typically narrow the scope by focusing on specific glyph or chart types. 

\bpstart{Mark Detection}~Many approaches assume that input visualizations are in a raster image format, where object detection is essential. For example, GlyphCreator \cite{ying_glyphcreator_2022} focuses on circular glyphs, and uses deep learning to perform object and bounding box detection. Similarly, visual elements in timeline infographics can be identified using deep learning \cite{chen_towards_2020}. OCR is typically used to recognize text elements \cite{poco_reverse-engineering_2017}. Since our input format is SVG, mark detection is not necessary.

\bpstart{Axis and Legend Detection}~Simple heuristics \cite{shukla_recognition_2008,poco_extracting_2018} or supervised learning \cite{poco_reverse-engineering_2017} can be used to extract
axes and legends.
However, these methods can still be error-prone. Since it is relatively easy to indicate where the axes and legends are, some tools expect users to provide such information \cite{poco_extracting_2018}. Mystique uses heuristics to find axes and legends, and provides a user interface for authors to correct potential mistakes.

\bpstart{Data Extraction} ~Previous work also addressed extracting data values from visualization images \cite{savva_revision_2011,jung_chartsense_2017} or vector graphics \cite{harper_converting_2017,harper_deconstructing_2014, masson2023chartdetective}. In Mystique, we demonstrate that a chart can be effectively reused without recovering the original data. Thus, data extraction is not necessary. 

\bpstart{Extraction of Visual Encoding and Spatial Arrangements} Inferring a visual encoding concerns the identification of relevant visual channel, data type, and potentially scale type. For glyphs with regular shapes (e.g., rectangles), visual encodings can be inferred using heuristics by combining information from mark type and axis \cite{poco_reverse-engineering_2017}. For custom glyphs (e.g., those used in infographics), \revise{sometimes the positions are not strictly encoded by data, but instead determined by specific spatial relationships or constraints. In these cases, current approaches usually classify charts into a predefined set of spatial arrangements~\cite{cui_mixed-initiative_2022,chen_towards_2020}. In Mystique, we break down the spatial arrangement of a chart into semantic components to handle more complex layouts.}


\bpstart{Chart Type Classification} Previous work also tackled the chart type classification problem. Most approaches are based on a simple chart taxonomy that roughly corresponds to mark types. For example, Revision \cite{savva_revision_2011} classifies chart images into 10 categories using SVM: area, bar, line, map, Pareto, pie, radar, scatter plot, table, and Venn diagram. This taxonomy is used in subsequent neural network-based methods \cite{poco_reverse-engineering_2017,jung_chartsense_2017}. In this work, we decided not to classify mark or chart types because such taxonomies are inadequate to capture the richness and variations of visualization design. Instead, we deconstruct charts into finer-grained semantic components.
\section{Overall Approach and Usage Scenario}
\label{sec:overview}

\subsection{Challenges and Processing Pipeline}


\begin{figure*}[!b]
\centering
\includegraphics[width=\linewidth]{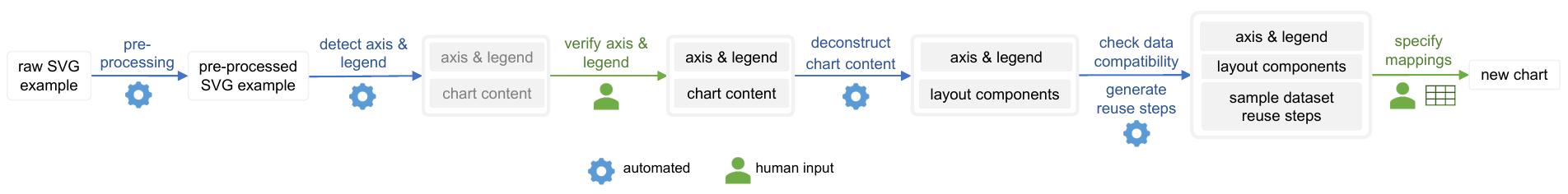}
\caption{The end-to-end pipeline for reusing an SVG chart to create a new chart in Mystique.}
\label{fig:overview}
\end{figure*}

\markup{We identify two main challenges in understanding and reusing SVG chart layouts. First, the semantic information such as mark attributes and hierarchical grouping in SVG specifications are not reliable. As we examined online SVG charts from different sources, we found that a majority of them were not readily usable. The following observations characterize the uncertainties in semantic structures: \textbf{ (1) Inconsistent SVG Element Types}: the same mark type may be represented using different types of SVG elements. For instance, we have observed in multiple examples that a rectangle mark is represented using a $<$path$>$ element, and an axis is drawn as a thin rectangle. Thus, we cannot determine the mark type based on the SVG element type; \textbf{(2) Missing Absolute Positions}: absolute positions of elements in a chart are crucial for determining their graphical roles and spatial relationships. But in many cases, an element's position is not expressed in absolute coordinates. Instead, the positions are often described using transformations such as ``translate'' or the matrix function; \textbf{(3) Noise in Scene Graph}: it is not trivial to distinguish visualization marks from graphical objects that are not part of the main visualization, which include off-screen tooltips, transparent or white rectangles serving as backgrounds, and random watermarks drawn with $<$path$>$ elements; and \textbf{ (4) Arbitrary Grouping of Elements}: 
the grouping of elements can be unpredictable. For example, grouped bar charts created using different tools exhibit vastly different grouping structures, and axis labels can be either within one group or in their own individual groups~\cite{chen2023state}.} 

\markup{Second, the \textbf{schematic congruency} \cite{satyanarayan_critical_2019}  between user's data and an example's layout structure is not guaranteed: the data may be formatted or structured in a way that cannot be readily applied to an extracted layout. To reuse the layout, users may face two obstacles: conceptualizing the
expected data layout and implementing the transformation \cite{wang_falx_2021}. Previous work (e.g., Falx \cite{wang_falx_2021}) has demonstrated the possibility of using program synthesis to automatically infer the required layout and transform the input data. Such an approach removes the need to perform data transformation, but
still requires the data to be in a tidy format \cite{wickham_tidy_2014}. It is thus likely only
going to work on input data that can be automatically morphed using
the predefined data transformation operations in the system. If the
system cannot find a feasible data transformation process (e.g., when
the input data is not in a tidy format), users would have no clue what
went wrong and how to intervene.}

\markup{To address these two challenges, Mystique adopts a pipeline (\cref{fig:overview}) for extracting and reusing layouts. 
The pipeline consists of the following stages:
\textit{pre-processing} raw SVGs~(\cref{sec:svg})}, \textit{detecting} axis and legend information with a user \textit{correcting} axis and legend detection mistakes~(\cref{sec:axisLegend}), \textit{deconstructing} the chart content into semantic components \revise{that jointly determine the chart layout~(\cref{sec:chartDecomp}), 
\textit{generating} chart reuse steps after checking data compatibility}~(\cref{sec:reuseUI}), and finally the user \textit{specifying} how these components map to data to create a visualization with new data. 
While the user collaborates with the system throughout the process, Mystique strives to minimize their effort and required skills. 
\markup{The pre-processing stage handles inconsistent element types, missing absolute positions, and noise in SVG specifications; the mixed-initiative stages of axis \& legend detection and chart deconstruction handle arbitrary grouping of elements; and the pipeline includes a step to help users understand expected data layout through auto-generated sample data and compatibility checking. 
Users can get a clear idea of what the input dataset should look like, and prepare the data either from scratch or by transforming an existing dataset. Mystique does not directly address the issue of implementing the transformation. By treating data preparation as a separate stage, Mystique can be used in conjunction with interactive data transformation tools like Data Wrangler \cite{kandel_wrangler_2011}, Tableau Data Prep \cite{tableau_prep}, and Trifacta \cite{trifacta}, which can support a wide range of idiosyncratic input data.}  

\subsection{Usage Scenario}


In this section, we illustrate the pipeline and how a user interacts with Mystique using a treemap grouped bar chart~\cite{treemapBar} as an example. The complete reuse process is presented on our demo website, \url{https://mystique-vis.github.io}. The chart uses a hybrid layout design, where the overall grouped bar chart representation shows trade values in different years, with the bar height encoding the total value. Within each bar, a two-level treemap shows the proportion contributed by each country, colored and grouped by continent. 
Currently, the only way to create \revise{a chart of such bespoke layout} is to program using libraries like D3, and it requires deep D3 expertise and significant time to modify the code for a new dataset. In contrast, Mystique enables the reuse of such chart on new datasets with simple interactions. Mystique does not require the user to learn a new language or framework. It only expects the ability to understand the chart.

\begin{figure*}
  \centering
  \begin{minipage}[b]{.33\textwidth}
    \subcaptionbox{
    Result panel for axis \& legend detection. Errors can be fixed through simple interactions~(e.g., in the figure the user is dragging ``1985'' from the chart to a higher-level label box).
    \label{fig:legend_axis_UI}}
      {\includegraphics[width=\textwidth, height=0.4\textwidth]{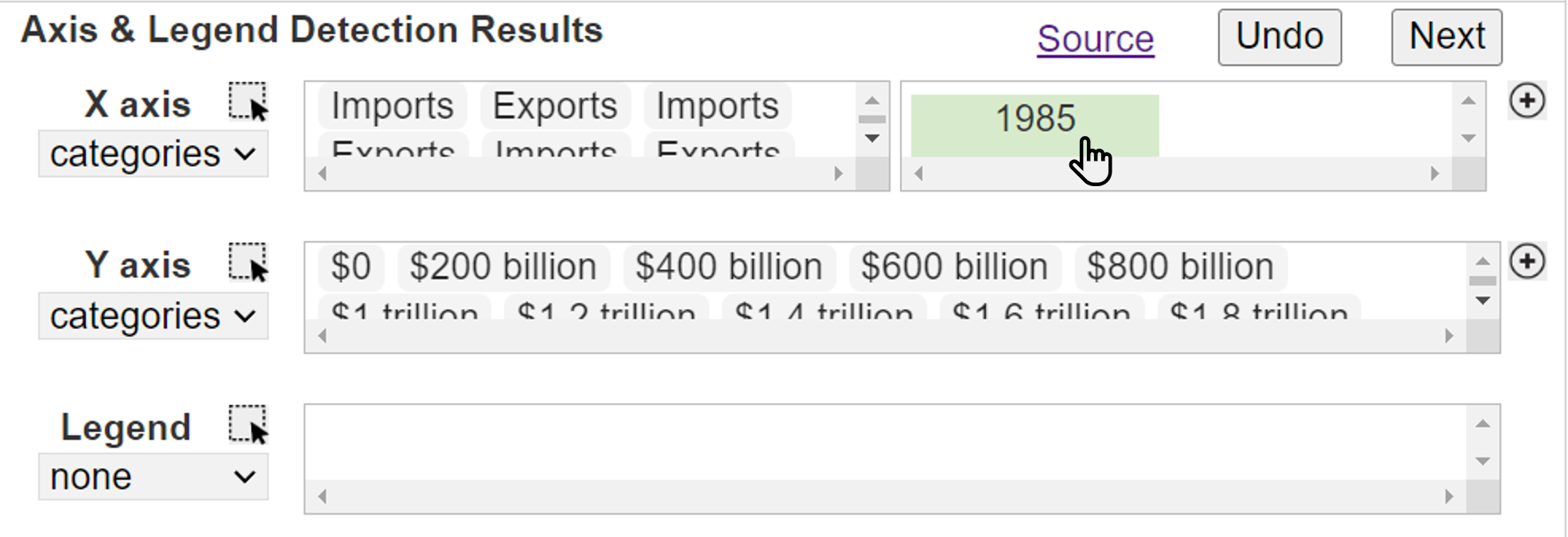}}
    \subcaptionbox{Sample data generated by Mystique for the treemap grouped bar chart example. \label{fig:sample_data}}
      {\includegraphics[width=\textwidth]{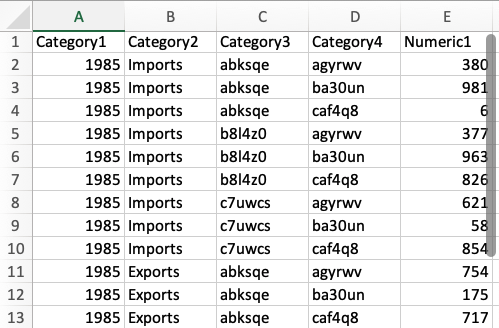}}%
  \end{minipage}%
  \hfill
  \begin{minipage}[t]{.66\textwidth}
    \subcaptionbox{Mystique's reuse UI consists of 1) Canvas, 2) Reference Panel, 3) Dataset Compatibility Panel, 4) Step Indicator, 5) Instruction Panel, and 6) Data Table Panel. \label{fig:reuseUI}}
      {\includegraphics[width=\textwidth]{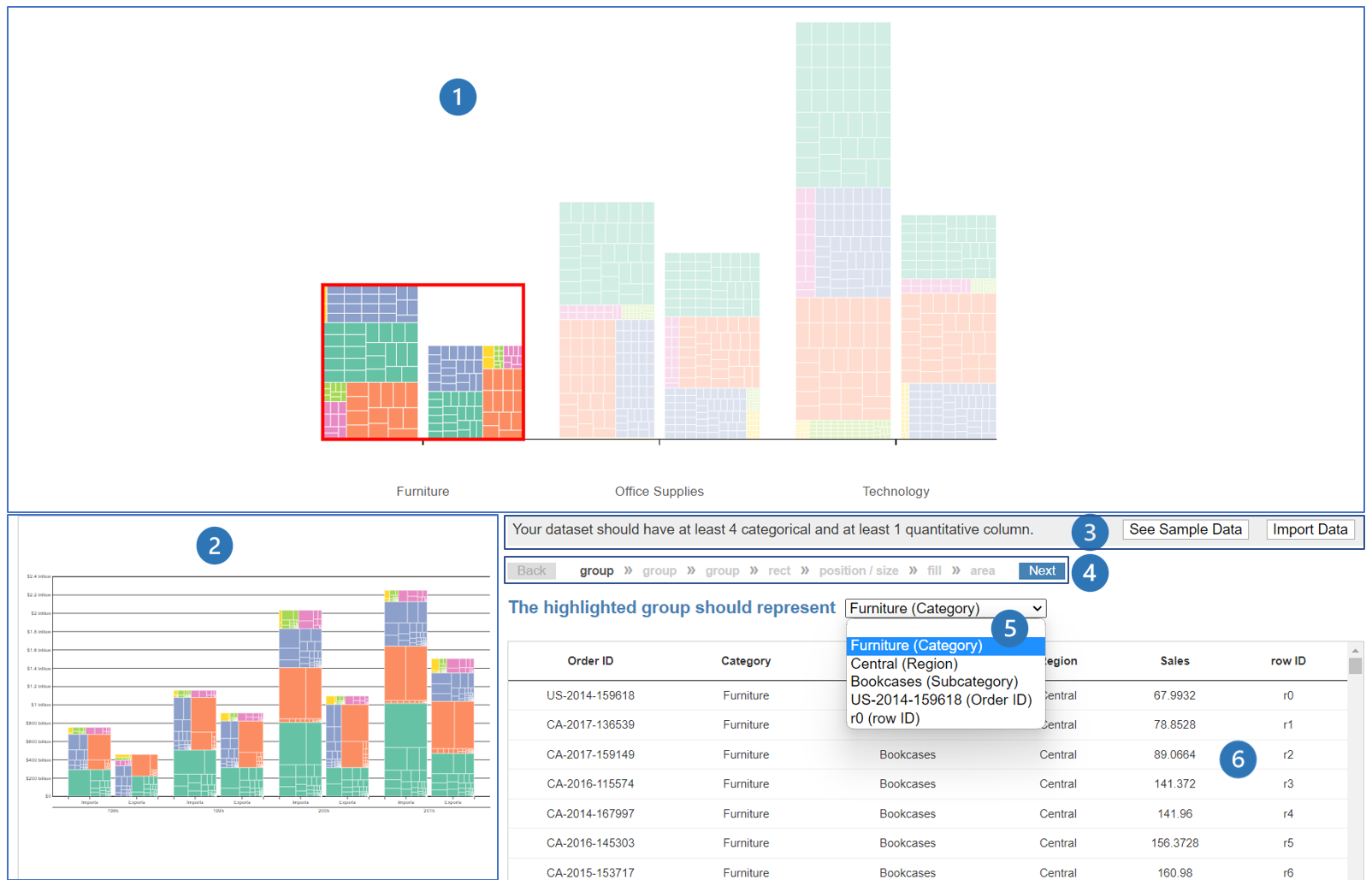}}%
  \end{minipage}%
  \caption
    {%
      (a) result panel for axis\& legend detection; (b) a sample dataset provided by Mystique; (c) the reuse UI consisting of six components.
      \label{fig:axisLegendReuseUI}%
    }%
\end{figure*}

With the treemap bar charts loaded, the user sees the automatically detected axis and legend information including labels and data types in the result panel~(\cref{fig:legend_axis_UI}). Since Mystique does not guarantee 100\% accuracy, it allows the user to fix potential detection errors through simple interactions. Once the user has verified the detection results, Mystique analyzes the semantic structure of the main chart content, synthesizes data requirements as well as chart reuse steps, and then prepares a wizard interface that guides the users to reuse the example.


\begin{figure*}[!b]
  \centering
      {\includegraphics[width=0.2\textwidth]{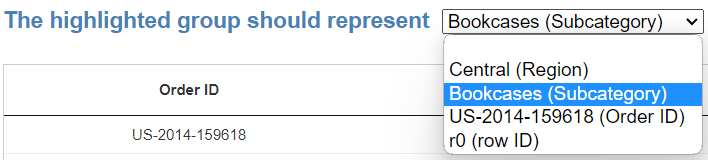}}\hspace{20px}\vspace{5px}
      {\includegraphics[width=0.2\textwidth]{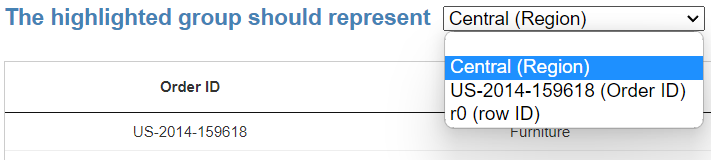}}\hspace{20px}
      {\includegraphics[width=0.2\textwidth]{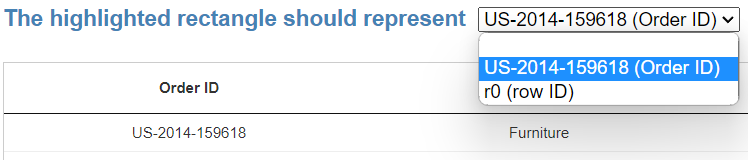}}\hspace{20px}
      {\includegraphics[width=0.2\textwidth]{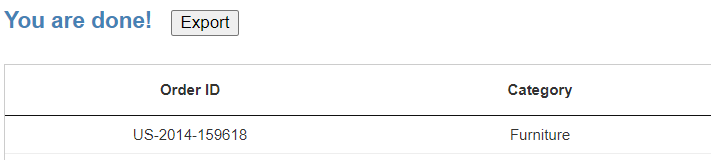}}
    \subcaptionbox{ \label{fig:step2}}
      {\includegraphics[width=0.2\textwidth]{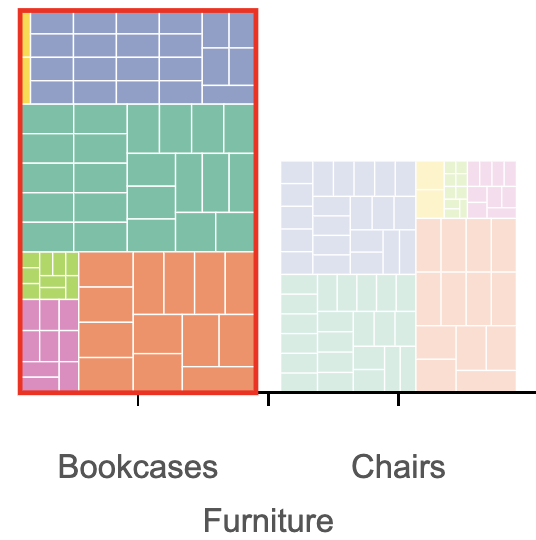}}\hspace{20px}
    \subcaptionbox{ \label{fig:step3}}
      {\includegraphics[width=0.2\textwidth]{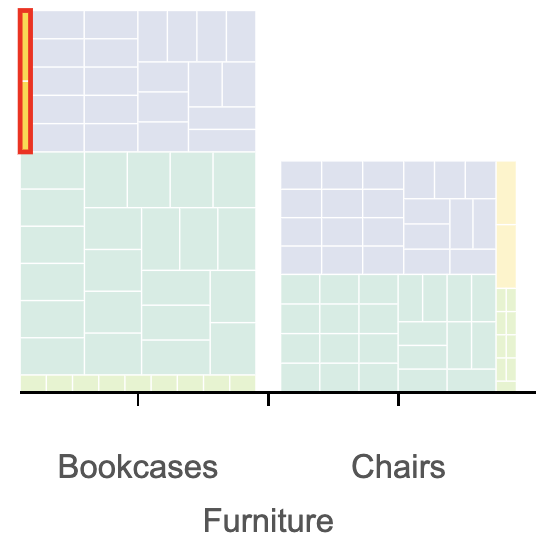}}\hspace{20px}
    \subcaptionbox{ \label{fig:step4}}
      {\includegraphics[width=0.2\textwidth]{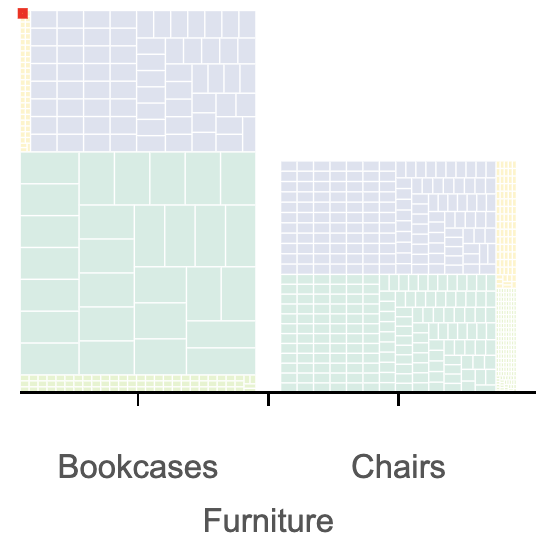}}\hspace{20px}
    \subcaptionbox{ \label{fig:step8}}
      {\includegraphics[width=0.2\textwidth]{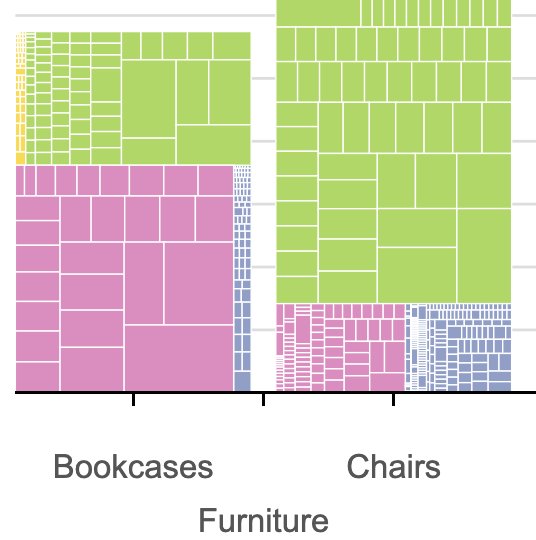}}
  \caption
    {%
      (a) subgroups mapped to ``Subcategory''; (b) color-based groups mapped to ``Region''; (c) rectangles mapped to ``Order ID''; (d) the final chart with all encodings applied. To save space, we only show the part of the chart representing \textit{Furniture}~(the remaining part is updating similarly), and (d) is further cropped vertically to more clearly show the changes in rectangle size and layout.
      \label{fig:reuseSteps}%
    }%
\end{figure*}

\definecolor{blue11}{RGB}{46, 117, 182}
\newcommand*\bluecircle[1]{\tikz[baseline=(char.base)]{
\node[shape=circle,draw,inner sep=0pt, blue11,fill=blue11,text=white] (char) { {#1}}
}}

The user first sees the original chart displayed at the bottom left as a reference (\cref{fig:reuseUI} \bluecircle{2}).  The Canvas (\cref{fig:reuseUI} \bluecircle{1}) initially shows the same chart without any axis or legend information, so that the visual representation can be updated with new data. The interface also shows a guideline on the minimum number of categorical and quantitative data columns required to reuse the example (\cref{fig:reuseUI} \bluecircle{3}). The user can download a sample dataset~(\cref{fig:sample_data}) which helps them understand the expected data format. After the user uploads their own dataset about product sales, which is shown in the Data Table Panel (\cref{fig:reuseUI} \bluecircle{6}), Mystique generates an ordered set of steps to reuse the chart in the Step Indicator (\cref{fig:reuseUI} \bluecircle{4}), and highlights the steps that have been completed so far. At each step, the Canvas highlights different parts of the visualization to solicit user specification (\cref{fig:reuseUI} \bluecircle{1}), and the user chooses the visual channel and data field for mapping through drop-down menus (\cref{fig:reuseUI} \bluecircle{5}). The visualization updates based on the user's operation at every step. The user can always go back to previous steps by clicking the Back button in the Step Indicator.

Mystique starts by asking what the highlighted highest-level group in the Canvas should represent in the new dataset~(\cref{fig:reuseUI} \bluecircle{5}): everything but this group fades into the background. The user selects a value from the ``Category'' attribute, resulting in three highest-level groups with labels from the ``Category'' attribute shown in~\cref{fig:reuseUI} \bluecircle{1}. The user clicks on the ``Next'' button in the Step Indicator to proceed. \revise{Mystique then highlights the first treemap bar and asks what this subgroup should represent~(\cref{fig:step2}); similar to the previous step, the user selects a value from the ``Subcategory'' attribute, which leads to two subgroups updated with labels~(\textit{Bookcases} and \textit{Chairs}) from the ``Subcategory'' attribute.
Mystique next highlights the yellow-colored group of rectangles in the first treemap bar and asks what it should represent~(\cref{fig:step3}); the user selects a value from the ``Region'' attribute, resulting in four packed groups of rectangles~(previously there are six) of distinct colors that correspond to four regions within each product subcategory.
In the following step, Mystique highlights the first rectangle and asks what it should represent~(\cref{fig:step4}); the user selects a value from the ``Order ID'' attribute, resulting in the system populating rectangles, one for each distinct order ID.}
After that, Mystique starts handling encodings. Mystique first guesses that the height of each treemap bar encodes ``Sales'', which is accepted by the user. Mystique continues to infer that the ``fill'' channel of each rectangle is mapped to ``Sales,'' which is undesired, thus the user changes `Sales'' to ``Regions.'' Mystique further suggests that the ``area'' channel of each rectangle is mapped to `Sales,'' 
resulting in the final visualization shown in full opacity~(\cref{fig:step8}). The user can export the chart and import it into Data Illustrator~\cite{liu_data_2018} for further customization such as axis range and color scheme. 

\section{SVG Pre-processing}\label{sec:svg}


Since real-world SVG charts are messy and far from ready to be reused, we perform the following preprocessing tasks to clean up a given chart. 

\bpstart{Converting to JSON Representation} The XML format used in the SVG charts is not very amenable to processing and inference. We thus first 
convert the given chart into a JSON object, recording attribute-value pairs for each element.
We also record additional information such as each element's parent. In cases marks' visual styles (e.g., stroke color) are stored in their parent groups, we record these styles as the marks' attributes.

\bpstart{Obtaining Absolute Position} 
To calculate the absolute position of each element, our SVG parser applies an element's transformation~(if any) to its children elements.
The transformations can accumulate across multiple levels 
and the parser will iteratively update the transformation until reaching a leaf node.

\bpstart{Identifying and Filtering Rectangle Marks} 
To capture rectangle marks represented as $<$path$>$ elements, 
we run 
a rectangle test for each $<$path$>$ element. This test parses the \textit{d} attribute of the $<$path$>$ element, records the vertices on this path, and determines whether the vertices are forming four corners of a rectangle. A $<$path$>$ element that passes this test is converted into a rectangle element in the JSON object, and its attributes are updated accordingly. We perform similar tests for lines represented as paths. Our parser also excludes dummy rectangles, such as transparent rectangles and rectangles with zero height and width.

\section{Axis and Legend Detection}
\label{sec:axisLegend}

We develop heuristics to detect axes and legend in a chart based on the relative positioning of marks, texts, and lines. For example, an axis area typically consists of a set of text labels, a set of small ticks close to corresponding labels, and an axis line spanning the vertical or horizontal range of the ticks; a legend area can be either a set of horizontal or vertical [mark, text] pairs~(the discrete case) or a gradient-colored bar associated with ticks and numbers~(the continuous case). \markup{The accuracy of the heuristics is reported in~\cref{quantatitive}, and the supplementary materials\footnote{All the supplementary materials are available at \url{https://osf.io/pt3yq/}} contains more details}.

The heuristics don't guarantee 100\% accuracy, and thus we build a user interface in Mystique to allow fixing potential errors through simple interactions.
The user interface consists of two areas: the chart area displaying the original SVG example, and the result panel under the chart area (\cref{fig:legend_axis_UI}). The result panel includes three sub-panels: x-axis, y-axis, and legend, showing the extracted labels, respectively.
The background colors of legend labels indicate the color mappings extracted from the legend. Mystique also infers the data types for the x- and y-axis, which are displayed in the form of a drop-down menu. 

Five kinds of detection mistakes have been observed: missing some labels~(M1), false-positive axis labels~(M2), missing higher-level labels~(M3), missing axis or legend~(M4), false-positive axis or legend~(M5). To enable users to fix these mistakes, we enable the following features: 
(1) drag-and-drop over chart texts into or out of the result sub-panels~(M1--3); (2) the \includegraphics[height=\fontcharht\font`\B]{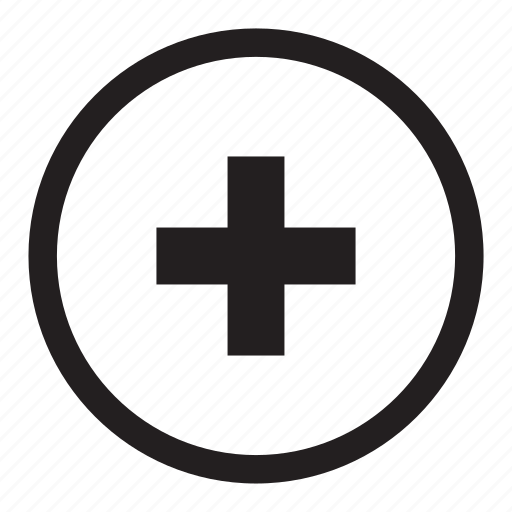} buttons right to the label display boxes that add display boxes for higher-level labels~(M3); (3) the \includegraphics[height=\fontcharht\font`\B]{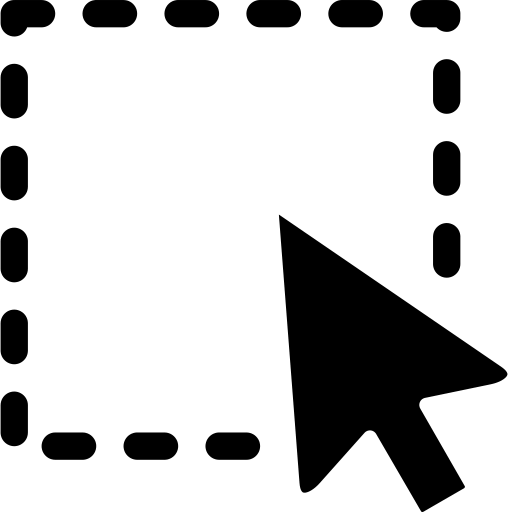} buttons left to the label display boxes to activate a region select tool for users to select an area that covers the missing axis or legend~(M4); (4) the drop-down selections left to the label display boxes, which allow users to remove false-positive axis or legend~(set to ``none'') and modify axis label data type~(M5).

\section{Deconstructing Chart \revise{Layout}}
\label{sec:chartDecomp}
After axes and legend are correctly identified, Mystique removes them from the scene graph and deconstructs the main chart content. 

\subsection{Semantic Components: \revise{GREC}}
\label{sec:GLEC}

To understand and decompose a chart \revise{layout}, one potential way is to classify the chart based on a fine-grained taxonomy with categories. We decided not to take this approach for three reasons. First, a chart might not clearly fit into a predefined type; for example, the chart in \cref{fig:reuseUI} \bluecircle{2} integrates elements from a treemap and a grouped bar chart into a single design. Second, charts can have nested structures~(e.g., small multiples) and thus require deconstruction into multiple instances of the same chart type. Finally, knowing the fine-grained chart type is still not enough to reuse \revise{its layouts}; for instance, given a stacked bar chart, we still need to obtain information such as the orientation of the stacked relationship and the distance between the groups. 
Therefore, we decided to deconstruct a chart into the following four types of semantic components \revise{that jointly reflect its layout}: \textit{grouping} (G), \textit{spatial relationships} (R), \textit{encodings} (E), and \textit{graphical constraints} (C).

\bstart{Grouping (G)} refers to the hierarchical clustering of rectangles that reflects the semantic structure of the visualization. We make a further distinction between two kinds of groups: \textit{collection} and \textit{glyph}. Rectangles in a group in the diverging stacked bar chart (\cref{fig:teaser}d) are placed in a horizontal stack relationship, and they represent different data cases; in contrast, rectangles in a group in the bullet chart (\cref{fig:teaser}c) are placed according to some graphical constraints (i.e., left and middle aligned) and they represent the same data case. We refer to the former type of group a \textit{collection} and the latter a \textit{glyph}, based on the unified terminology used in recent visualization authoring tools \cite{satyanarayan_critical_2019} and the visualization object model used in the \markup{Mascot} grammar \cite{liu_atlas_2021}. 

\begin{figure}[ht]
\centering
\vspace{-2mm}
\includegraphics[width=\linewidth]{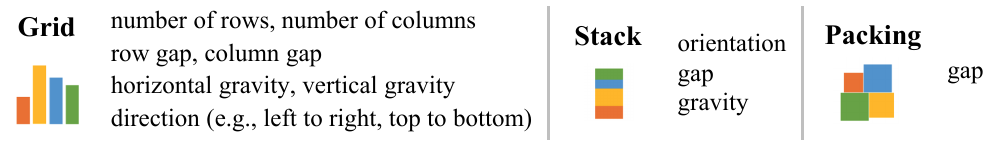}
\caption{Three types of spatial relationships (grid, stack, and packing) and their parameters in Mystique. 
}
\label{fig:layouts}
\end{figure}

\bstart{Spatial Relationships (R)} estimate the relative placement and organization of same-level rectangles or groups. Mystique currently supports three types of relationships: grid, stack, and packing~(\cref{fig:layouts}).

\noindent \textbf{Encodings (E)} specify the mapping between data attributes and visual properties of rectangles or groups.

\bstart{Graphical Constraints (C)} enforce requirements (e.g., data-related alignment) on the spatial arrangements of rectangles or groups. For example, in \cref{fig:teaser}d, all the gray rectangles representing ``Neither agree nor disagree'' are aligned in the center. The primary difference between a graphical constraint~(C) and a spatial relationship~(R) is that the former can transcend groups and be applied to selected marks only, while the latter computes the positions of all the children in a group using an iterative algorithm. In certain cases, aspects of a spatial relationship may be similar to graphical constraints. For instance, in \cref{fig:teaser}e, multiple collections of vertically stacked bars are organized as nested groups in a grid relationship, the contents of which are aligned by the bottom. Mystique treats such an alignment as a relationship parameter (called ``gravity''), not as a constraint.

Consider \revise{the layout of} the treemap grouped bar chart in 
\cref{fig:reuseUI}. Excluding the axes and grid lines, the main chart area can be described as 4 high-level \textit{groups} (G), each representing a year, arranged in a \textit{grid relationship with 1 row and 4 columns} (R). Within each group, two \textit{sub-groups} (G) of rectangles, representing Imports and Exports, are arranged in a \textit{grid relationship with 1 row and 2 columns} (R). Within each subgroup, six sub-subgroups (G) in distinct colors are arranged in a \textit{packing relationship} (R), each of which is composed of rectangles arranged in a \textit{packing relationship} (R). The \textit{height} of the bars representing Imports or Exports \textit{encodes} the \textit{total trade value} (E), the \textit{area} of the rectangles \textit{encodes} the \textit{trade value} (E) for a certain country, and the \textit{color} of the rectangles \textit{encodes} the \textit{continent} (E). In this chart, there are no spatial graphical constraints.
This GREC model is based on the \markup{Mascot} visualization framework \cite{liu_atlas_2021}, and the four components correspond to the outputs of four grammatical rules:  glyph generation, graphics-data join, visual encoding, and spatial arrangement.

\subsection{Group and Spatial Relationship Detection}
\label{sec:decompApproach}
\revise{We decided to jointly detect groups and spatial relationships because the inference of a group and the spatial relationship~(if any) in a group are inextricably linked.}
On the one hand, a relationship would be an effective criterion to decide whether a set of rectangles form a group; on the other hand, uncertainties may be involved if we try to infer the relationship without a complete set of rectangles from a group.
For this task, we have considered multiple approaches. 

We experimented with an \markup{SVM model~\cite{noble2006support}} trying to predict if a gap between a pair of rectangles constitutes a boundary between groups, but the results were not satisfactory~(the results are included in the supplementary materials). 
Another candidate, the decomposition approach in Chartreuse~\cite{cui_mixed-initiative_2022}, is not directly applicable because this method was primarily designed for clustering customized infographics glyphs. It is also unclear if it works on charts with nested structures.

Our final solution is a customized bottom-up hierarchical clustering algorithm~\cite{nielsen2016hierarchical} to capture the semantic groups of rectangles.
Generally, hierarchical clustering starts by treating each object as an individual cluster, and combines pairs of clusters until one final cluster containing all objects is formed. Applying this approach to our problem, the clustering algorithm consists of two steps: inferring the lowest-level groups and spatial relationships based on pairwise information for the rectangle marks (\cref{sec:lowestLayout}), and iteratively merging groups to complete a nested structure (\cref{sec:groupMerging}).
In this process, two major components need to be determined: a \textit{distance function} $D$ that computes how closely related two rectangles or two groups are, and a \textit{linkage function} $K$ that merges objects into hierarchical clusters based on the distance information. We design these functions by taking into account the semantics of grouping in data visualization. 
\revise{
To avoid overfitting
and improve generalizability, we develop and tune the algorithm on a training chart set, and evaluate the algorithm on a separate test chart set (details
in~\cref{quantatitive}).}
The pseudo-code for the algorithm is included in the supplementary materials.

\begin{figure*}[ht]
\centering
\includegraphics[width=1.0\linewidth]{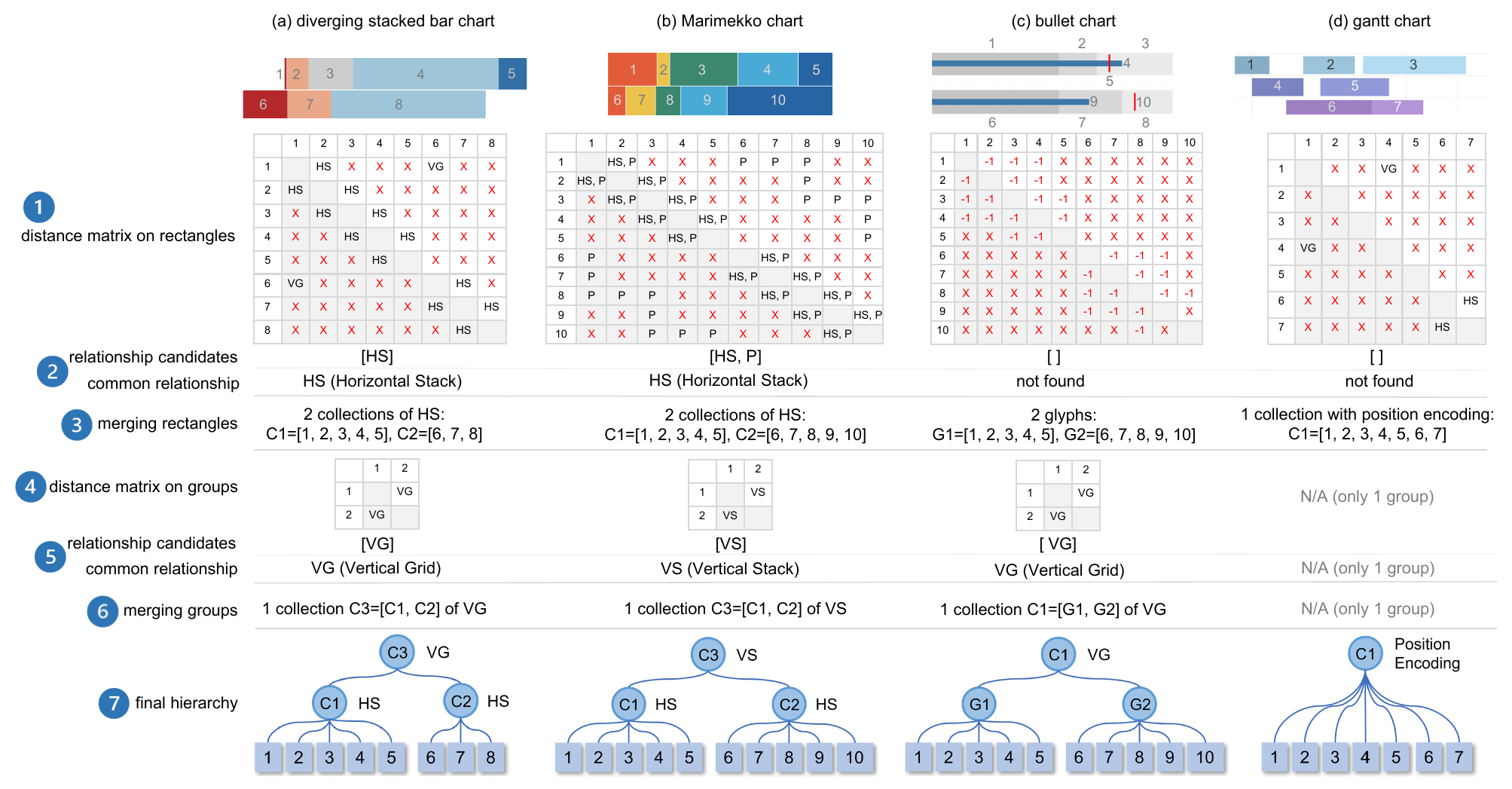}
\caption{\revise{Chart decomposition process for four different chart segments. The matrix cells store the results from the distance function for each pair of rectangles or groups. Since the matrix is symmetrical, the gray cells do not have to be computed. HS~(VS) stands for horizontal~(vertical) stack, HG~(VG) stands for horizontal~(vertical) \markup{grid}, P stands for packing, -1 means overlapping rectangles, and X means \textit{null}.}}
\label{fig:diagram1}
\vspace{-4mm}
\end{figure*}

\subsubsection{Lowest-Level Groups and Spatial Relationships Detection}\label{sec:lowestLayout}
The algorithm starts with detecting the lowest-level spatial relationships and groups. 
We use local spatial relationship information, instead of a numerical value only, to characterize how related two rectangles are. 

\bpstart{Distance Function}
We define a distance function $D$ that takes a pair of rectangles as input and outputs one or more relationship categories that can describe their spatial relationship. $D$ supports five relationship categories: stack (with orientation and gap), grid (with orientation and gap), packing (with gap), overlapping, and \textit{null}. 
We use the following criteria to decide between the stack, grid, and packing relationships: for all three, the rectangles should not overlap; grid and stack require that the union of the two rectangles' bounding boxes do not intersect with any other rectangle~(e.g., pairs~$(1,2)$ and~$(1,6)$ in \cref{fig:diagram1}a \bluecircle{1}), and packing is applicable when a universal gap parameter exists regardless of orientation~\markup{(i.e., the gap between any neighboring rectangles is constant)}. Thus, $D$ rules out the packing relationship once inconsistent gaps are observed (e.g., $(1,2)$ and $(1,4)$ in \cref{fig:diagram1}d \bluecircle{1}).


\bpstart{Linkage Function}~To merge rectangles into lowest-level groups, a linkage function $K$ computes the following:

\noindent\textit{1. Construct Distance Matrix} ~(\cref{fig:diagram1} \bluecircle{1}). $K$ first computes the relationship categories for every pair of rectangles using $D$, and stores the results in a matrix.
\revise{To accommodate noises of mark positioning in SVG charts and increase robustness, some small thresholds for relationship parameters~(\cref{fig:layouts}) are adopted~(details in the supplementary materials).}
\Cref{fig:diagram1} \bluecircle{1} illustrates the matrix results for four different charts. Since the matrix is symmetrical, we only need to compute fewer than half of the entries (shown as white cells). \revise{Cells with red labels indicate either overlapping rectangles (\textcolor{red}{-1}) or no applicable relationships (\textcolor{red}{X})}.

\noindent\textit{2. Extract Common Relationship}~(\cref{fig:diagram1} \bluecircle{2}). $K$ finds if a common spatial relationship exists in the distance matrix. For instance, HS (horizontal stack) appears in all rows in \cref{fig:diagram1}a \bluecircle{1}, which means that for every rectangle, there exists at least one other rectangle that can be paired with it to form a horizontal stack relationship. In cases like the Marimekko chart~(\cref{fig:diagram1}b) and a Pyramid chart, multiple relationship candidates are available (the latter case has both the grid and stack relationships in common) and lead to different final groups. 
To choose the common relationship, stack is given priority over grid \markup{or} packing. \markup{In all other cases only one candidate exists.}

\noindent\textit{3. Merge Rectangles into Low-level Groups}~(\cref{fig:diagram1} \bluecircle{3}). Given the common relationship $L$, $K$ randomly picks an initial pair of rectangles satisfying $L$ and merges them into a collection $C$. It then looks for other rectangles that can be added to $C$ without breaking $L$. This iterative clustering continues until no new rectangles can be found.
$K$ then proceeds to find other lowest-level collections until all rectangles are assigned, requiring that collections do not overlap with each other. 
In cases where no common relationship was found but overlapping rectangles exist~(\eg in \cref{fig:diagram1}c, rectangles 1-5 and rectangles 6-10 form two glyphs), $K$ uses the $-1$ values in the distance matrix to cluster overlapping rectangles to form lowest-level glyphs. 
At this point, since the lowest-level groups are complete, the algorithm calls the distance function $D$ again to examine the lowest-level groups to update and augment $L$ with more details like gravity.
If no common spatial relationship was extracted in the second step and no glyph was found, or the requirement of non-overlapping groups was violated, $L$ groups all rectangles together and infers that the position of each rectangle encodes data (\cref{fig:diagram1}d).

\subsubsection{Iterative Group Merging}\label{sec:groupMerging}
After all the lowest-level groups are found, we start recursively merging these groups into higher-level groups by checking potential higher-level relationships between groups. 
Similar to \cref{sec:lowestLayout}, the algorithm first constructs a distance matrix recording potential relationship information for all possible group pairs using $D_g$~(\cref{fig:diagram1} \bluecircle{4}), extracts a common relationship from the matrix~(\cref{fig:diagram1} \bluecircle{5}), and finally merges the groups into higher-level groups using $K_g$~(\cref{fig:diagram1} \bluecircle{6}). $D_g$ and $K_g$ share the same logic with $D$ and $K$, respectively, where the only difference is the input objects.
We recursively apply this process until ending up with a single collection of groups containing all rectangles or termination during this process~(i.e., no relationship is found or collections of groups overlap). 

For example, in both \cref{fig:diagram1}a and~\cref{fig:diagram1}c, the distance matrix for groups is $\big(\begin{smallmatrix}
  null & VG\\
  VG & null
\end{smallmatrix}\big)$; thus a higher-level vertical grid relationship is found, and two lowest-level groups are merged into one. In \cref{fig:diagram1}b, the distance matrix for groups is $\big(\begin{smallmatrix}
  null & VS\\
  VS & null
\end{smallmatrix}\big)$, leading to a higher-level group with a vertical stack relationship. In \cref{fig:diagram1}d, all rectangles are encoding-based and clustered into one group; thus there is no need to proceed to higher-level handling. The final decomposed hierarchies are shown in \cref{fig:diagram1}~\bluecircle{7}.

\subsection{Encodings Inference\label{sec:encInfer}}


\noindent \textbf{Encodings for Groups.} In cases where the iterative group merging described in \cref{sec:groupMerging} aborts~(which means several groups sharing the same semantic structure are found but they cannot be merged further to form higher-level hierarchies), Mystique collects these groups to make the final grouping result and marks their $x$ and $y$ positions as encoded with data; e.g., the small-multiples bar chart~(\cref{fig:teaser}g).

\bpstart{Encodings for Rectangles} Mystique considers six visual channels of a rectangle mark in encoding detection: $x$, $y$, $width$, $height$, $fill$, and $area$. \Cref{table:encoding-inference} summarizes our inference rules. When the lowest-level groups are glyphs, Mystique applies the rules for width/height and x/y to extract encodings for each set of corresponding rectangles~(e.g., four bars of the same color across groups in \cref{fig:teaser}c).

\begingroup
\renewcommand{\arraystretch}{1.1}
\begin{table}[ht]
\caption{Each visual channel is considered to be encoding data if the corresponding condition is met.}
\scriptsize
\begin{tabular*}{\linewidth}{lp{.36\textwidth}}
\toprule
\textbf{Channel}       &      \textbf{Condition}        \\ \toprule
fill            &     rectangles in the chart content have different fill colors                 \\ \midrule
area      &            lowest-level spatial relationship is packing               \\ \midrule
width/height         &       lowest-level spatial relationship is grid or stack, and rectangles have varying widths/heights                   \\ \midrule
x/y      &           lowest-level spatial relationship is a one-directional grid without the gravity parameter
                  \\ \bottomrule
\end{tabular*}
\label{table:encoding-inference}
\end{table}
\endgroup

\subsection{Graphical Constraints Detection}
Information regarding graphical constraints is recorded during the chart deconstruction.
Mystique currently checks and records two kinds of graphical constraints: (1) the alignment constraint within a glyph (e.g., the gray and blue rectangles in a glyph are left and middle aligned in \cref{fig:diagram1}c) and (2) any customized alignment of stacked bars in a grid relationship~(e.g., \cref{fig:teaser}d). The former is important when rendering a new glyph-based visualization in the reuse UI because the alignment constraints within glyph groups are not automatically enforced by spatial relationships. The latter is usually data-dependent; for instance, the stacked bar groups in \cref{fig:diagram1}a are aligned so that all the rectangles representing ``Neither agree nor disagree'' share the same center coordinate. Since such information have to be provided by the user with a new dataset, Mystique only records the alignment constraint based on the fill color property, and asks the user to provide data-related constraint specification during the reuse stage. 

\section{\revise{Data Compatibility and Step Generation}}\label{sec:reuseUI}

\subsection{\revise{Data Schema Inference}}\label{sec:checkingData}
To mitigate the risk of example misuse, Mystique infers data schema from the deconstructed semantic components. First, Mystique calculates $C_{group}$, the number of categorical data fields required to generate the grouping structure: each level in the deconstructed grouping structures corresponds to a unique categorical field. For instance, the diverging stacked bar chart in \cref{fig:teaser}d has a two-level nested structure 
which requires at least two categorical fields. 

Mystique also infers the number of categorical fields ($C_{encode}$) and the number of quantitative fields ($Q_{encode}$) required for encodings. The number of encoded channels is used as the number of data fields, and the data field types are obtained through the extracted and user-corrected axis and legend information. Since it is possible that the same field may be used to generate grouping structures and encode visual channels (e.g., \textit{response type} in \cref{fig:teaser}d), Mystique uses $\max(C_{group},\ C_{encode})$ as the minimum number of categorical fields.

Based on this analysis, Mystique generates a sample dataset for a given example (\cref{fig:sample_data}). It tries to find data values for each field based on axis and legend labels. If such data is not available, the field values are generated as random strings or numbers. The sample dataset is then created as the permutation of all the field values. Mystique also displays a guideline on the minimum numbers of categorical fields and quantitative fields in the Dataset Compatibility Panel in the reuse UI~(\cref{fig:reuseUI}), and checks if the dataset satisfies this requirement whenever a new dataset is imported. Mystique issues a popup-dialog warning whenever the requirement is not met. Since users may use the same field to encode multiple channels, Mystique does not enforce the data schema as a strict rule and users can dismiss the dialog.

\subsection{Generating Reuse Steps}\label{sec:generateReuseSteps}
Based on the deconstruction results, Mystique generates a sequence of data mapping steps that guide users toward the creation of a new chart. The steps are arranged in the following order: (1) mapping groups to categorical or date fields, from the highest level to the lowest level, (2) mapping rectangle marks to categorical or date fields, (3) choosing visual channels and data fields for size, position, area and fill encodings. 

As mentioned in \cref{sec:encInfer}, by default Mystique chooses x/y as the channels for position encodings, and width/height as the channels for size encodings. In some visualization designs, however, position and size encodings can be interchangeable and the distinction between the two may not be clear-cut. Consider the range chart in \cref{fig:teaser}b, it applies position encodings to the top segment and the bottom segment of each rectangle, which represent the daily maximum and minimum temperature respectively. However, it is also reasonable to infer the presence of a size encoding, where the height encodes the temperature range. In such ambiguous cases, Mystique is unable to clearly distinguish between position and size encodings. It thus provides multiple possible visual channels (e.g., \textit{top side}, \textit{bottom side}, and \textit{height}) through a drop-down menu for users to specify which channel should be used.

To update the visualization result at every step, we use \markup{Mascot.js (previously known as Atlas.js}~\cite{liu_atlas_2021}) as the underlying library. Its graphics-centric and procedural design enables displaying intermediate visualizations incrementally, so that users can evaluate whether they are on the right track.
The demo website showcases the results of reusing a variety of example charts, as a demonstration of Mystique's expressiveness.

\section{Evaluation}

\subsection{Evaluation of Deconstruction Algorithm}\label{quantatitive}

\textbf{Chart Corpus.}
A chart corpus is essential for developing and evaluating reuse algorithms and tools. 
Based on our scope described in \cref{sec:intro}, we first considered the rectangle-based charts in Beagle~\cite{battle2018beagle}. However, we found that its distribution over chart types is unbalanced: a majority of the charts are simple bar charts and histograms. 
Furthermore, the corpus contains charts created by only five tools, which is insufficient to capture the differences in SVG representations across different tools to achieve tool-agnostic reuse.
Therefore, we decided to collect our own corpus. Instead of prioritizing the quantity of charts, we sought to promote distribution balance and chart diversity in terms of layout, SVG representation (which is associated with the tool that produced the example), and visual style (e.g., colors theme and decoration). 

\begin{figure}[ht]
  \centering
  \includegraphics[width=.475\textwidth]{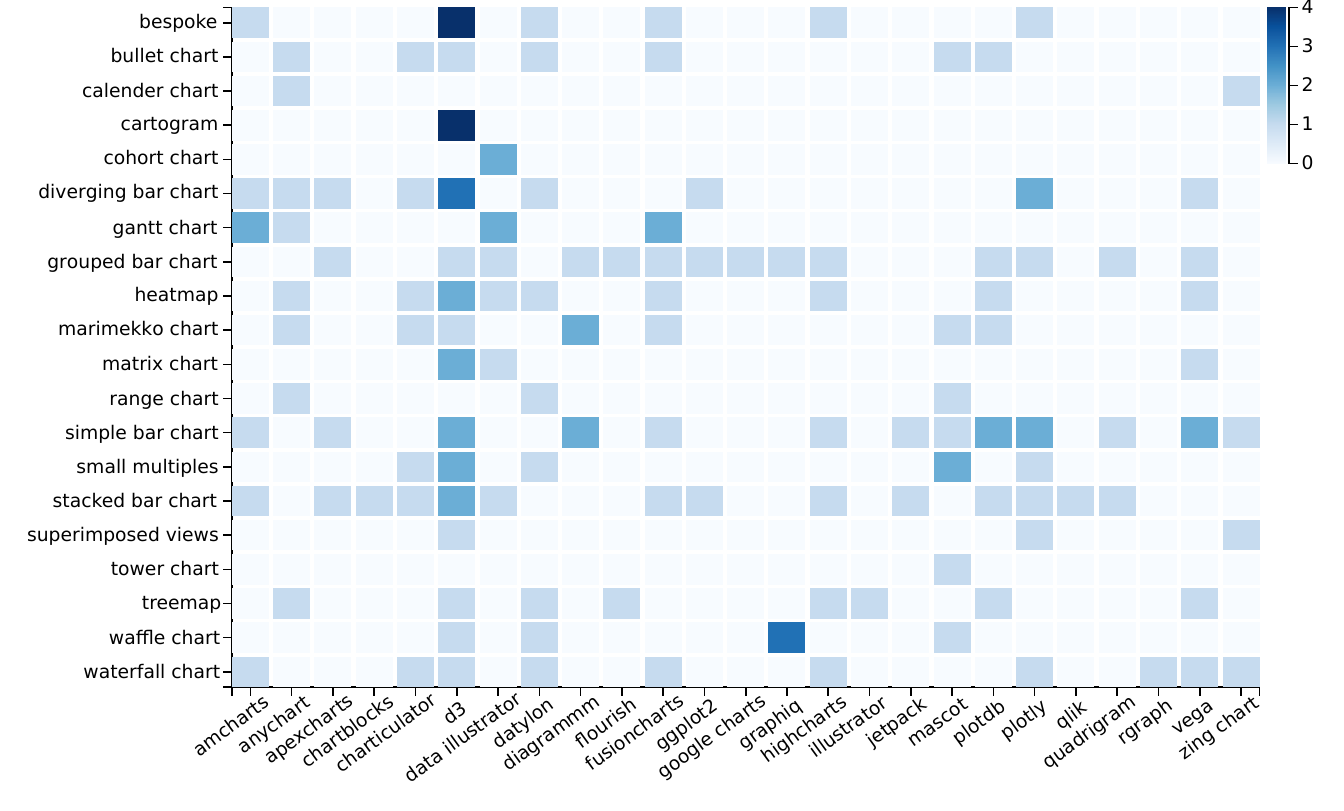}
  \caption{\markup{Distribution of the 150 examples  by chart design and tool. \label{fig:corpus}}}
  \vspace{-2mm}
\end{figure}

We started with the chart designs in Beagle (which were mostly created using D3), then referred to online catalogs such as the Chartmaker Directory \cite{chartmaker_dir}, which maintain comprehensive chart taxonomies and tools for each chart type. For each rectangle-based chart type in the catalogs, we sampled examples with design variations~(e.g., different orientations, alignments, and axis positions) produced by different tools. We used two methods to obtain the SVG representations of these examples. We downloaded the SVG elements, if available, and re-created examples using Charticulator~\cite{ren_charticulator_2018} and Data Illustrator~\cite{liu_data_2018} when examples are available only in demo videos or in raster images (e.g., Figma charts \& infographics \cite{figma_charts}). Finally, we browsed D3 galleries including bl.ocks.org~\cite{blockOrg} and Observable~\cite{observableplot} to identify and download bespoke designs that do not fall into predefined chart categories. 
The final collection, shown in~\cref{fig:corpus}, consists of 150 SVG charts produced by 25 tools, encompassing 17 chart types, 7 small multiple designs, \markup{3 superimposed views}, and 9 bespoke designs. 
The supplementary materials contain the 150 charts and figures showing details on the distributions of chart design in Beagle and our corpus.

\noindent \textbf{Axis \& Legend Detection Accuracy.}
On the 150 SVG charts, Mystique achieves $86.67\%$, $85.33\%$, and $90.67\%$ accuracy on the x-axis, y-axis, and legend inference, respectively. The supplementary materials contain detailed illustrations about the errors. \revise{All the detection errors can be corrected using the interactions described in \cref{sec:axisLegend}.} 

\begin{figure}[ht]
  \centering
  \includegraphics[width=.45\textwidth]{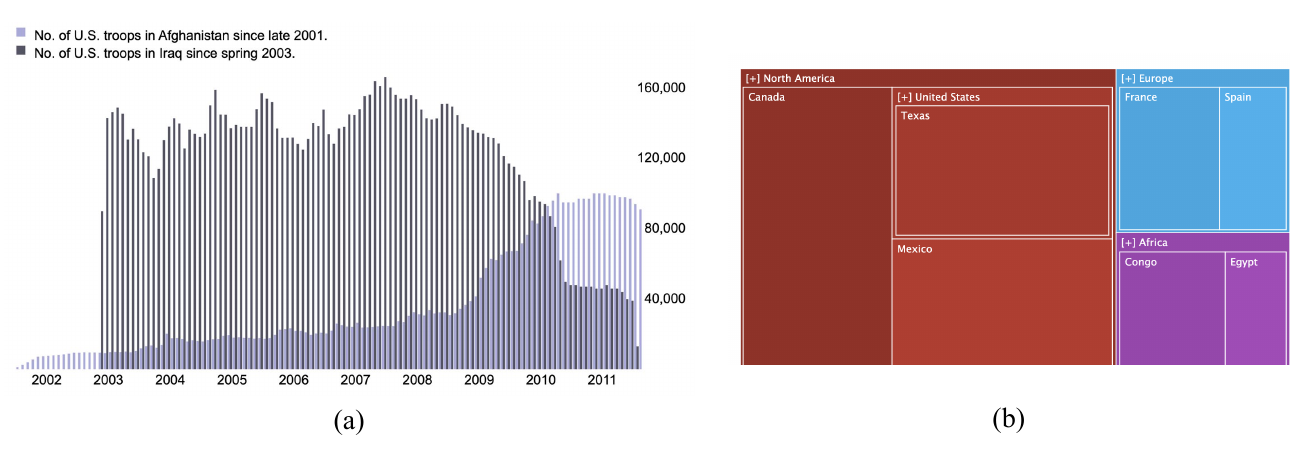}
  \caption{Mystique cannot detect overlapping groups: (a) two superimposed bar charts, (b) a treemap where \textit{country} rectangles are on top of \textit{continent} rectangles. \label{fig:errors}}
  \vspace{-2mm}
\end{figure}

\bpstart{GREC-based Chart Deconstruction Accuracy} Before developing the deconstruction algorithm (\cref{sec:chartDecomp}), we split our dataset into $105$ training and $45$ test charts~(a standard 7:3 ratio~\cite{goodfellow2016deep}) to avoid over-fitting and promote generalizability. We ensured that this 7:3 split is approximately maintained for both chart types and visualization tools.
We designed and fine-tuned our algorithm based only on the training set. 
The deconstruction algorithm achieves $96.19\%$~(= $101$/$105$) accuracy on the training set and $95.56\%$~(= $43$/$45$) accuracy on the test set.
Specifically, the algorithm fails in three cases where multiple charts~(e.g., the superimposed bar charts in \cref{fig:errors}a) or marks~(e.g., the treemap in \cref{fig:errors}b) are overlaid on top of one another.
Among the remaining three error cases, two represent rectangles using $<$line$>$ elements with large stroke-width values which is rarely seen; it also interferes with the cases where large-stroke-width $<$line$>$ elements represent axis lines, hindering our pre-processing method from accurately handling such cases. The remaining error case is due to relatively large gaps~($6$ pixels) between neighboring rectangles in a treemap---our threshold hyper-parameter for the gap parameter in a packing relationship is 5. These error cases are included in the supplementary materials.

\noindent \textbf{Algorithm Efficiency.} For each chart in the test set, we also recorded the time of main chart content decomposition. 
All the examples can be deconstructed within one second, except a heatmap created with Vega, which consists of more than 8K marks. The supplementary materials include detailed performance data.

\subsection{Chart Reproduction User Study}
To evaluate whether users can understand and follow the guidance from Mystique to reuse existing charts and produce new ones, we conducted a chart reproduction study~\cite{ren2018reflecting}.

\bpstart{Participants and Procedure}
We recruited $12$ participants~($5$ male, $7$ female) from the Washington metropolitan area. The statistics on how often they create data visualizations are as follows:
\markup{
Never~(1, $8.3\%$), A few times per year~(4, $33.3\%$), A few times per month~(5, $41.7\%$), and A few times per week~(2, $16.7\%$). The tools they have used are Excel~(6, $50.0\%$), Tableau~(6, $50.0\%$), D3.js~(5, $41.67\%$), ggplot2~(3, $25.0\%$), Vega/Vega-Lite~(2, $16.7\%$), Figma~(1, $8.3\%$), and Others~(4, $33.3\%$).
}

All the study sessions were conducted remotely on Zoom and each lasted about 1 hour. After a brief explanation of the study goal, we walked the participants through two sets of tutorials. The first set was on the UI for fixing axis and legend detection errors. The participants learned what kinds of errors could happen, and how to fix them through interactions. The second set of tutorials taught the participants about the workflows to reuse a simple bar chart and a grouped bar chart. The tutorials lasted about $25$ to $30$ minutes. The participants then were asked to complete four visualization creation tasks, reusing one chart for each task, with Mystique. At the end of the session, each participant completed a questionnaire, and we conducted a debriefing regarding their experiences using Mystique. Each participant was given a \$15 gift card as a token of appreciation.

\bpstart{Tasks}
We used a bullet chart (\cref{fig:teaser}c) for Task 1, a grouped stacked bar chart (\cref{fig:teaser}e) for Task 2, a diverging stacked bar chart (\cref{fig:teaser}d) for Task 3, and a range chart (\cref{fig:teaser}b) for Task 4. 
We chose these four charts to cover the major types of axis/legend detection errors: Task 2 requires adding a missing higher-level x-axis; Task 4 requires adding a missing y-axis; and Tasks 2 and 3 require changing incorrectly inferred field types. The four tasks also cover different semantic components with varying complexity in terms of nesting structures: Task 1 involves glyphs composed of multiple rectangles, whose positions are constrained; Task 2 has three levels of nesting; Task 3 involves the alignment constraint across collections; and in Task 4, the top and bottom segments of each rectangle encode data. 
For each task, we explained the input example chart, provided the participants with a new dataset, and described the schema and meaning of the new dataset. The participants were not shown any charts they need to create, only text descriptions of the target charts. 
Not to prime the participants or make the tasks too easy, we deliberately avoided mentioning visual channels in the instructions. For instance, the requirement for Task 4 is phrased as: ``\textit{Please create a visualization of the dataset. Each bar represents a day’s temperature range, from minimum to maximum temperature. The color represents the average temperature in the day}.''


\bpstart{Results}
All participants produced a visualization for all the tasks, but 5 out of 48 visualizations were not what the requirements asked for. The participants completed each task within five minutes on average, with Tasks 2 and 4 taking longer~(\cref{table:userStudy}): Task 2 involves three nesting levels, and Task 4 involves multiple combinations of the visual channel and data field.  
The participants rated their experience of using Mystique on a 5-point Likert scale (1: ``Strongly Disagree'' to 5:``Strongly Agree'') in the post-study questionnaire. The results are as follows: efficiency, $\mu=3.92$ $\sigma=1.04$; convenience regarding accommodating changes, $\mu=4.58$, $\sigma=0.49$; and \markup{comfort/confidence}, $\mu=4.25$, $\sigma=1.01$.

\begingroup
\renewcommand{\arraystretch}{1.1}
\begin{table}[ht]
\caption{The number of participants completing each task successfully, and the average completion time with standard deviation.}
\scriptsize
\begin{tabular*}{\linewidth}{cccc}
\toprule
Task       &      \# Successes  &     Average Time (minutes)   &      Standard Deviation         \\ \toprule
1            &     11            &  2.87     &  1.62                  \\ \midrule
2      &            11          &  4.35      &  2.47                \\ \midrule
3         &       11            &  2.86     &  2.40                  \\ \midrule
4      &           10          &  4.96      &  2.20                \\ \bottomrule
\end{tabular*}
\label{table:userStudy}
\end{table}
\endgroup

\vspace{-2mm}
We identified three main reasons why the participants failed to produce the correct visualization. First, some participants misunderstood the instructions. For instance, P3 produced a chart with three nested levels for Task 2, but the order of nesting was incorrect. He had trouble interpreting the instruction: ``\textit{The visualization should show the sales distribution for each region grouped by product category and subcategory}.'' Second, some participants did not understand the meanings of certain visual channels, particularly for Task 4. For instance, P4 stated that ``\textit{some sentences and words are not really clear for me. For example, there was a sentence where I can choose left-side, right-side [which were unclear].}'' Finally, one participant forgot that the visual channels were configurable too. After we reminded her, she could quickly create the desired chart during the debriefing.

Participants' experience in creating data visualizations affected their performance in completing tasks. Among the four participants who did not complete all tasks, three create visualizations ``a few times a year,'' and one ``never'' specifically created visualizations. It was harder for them to follow the instructions from Mystique. 

We also recorded the number of times participants clicked on the Back button to review previous steps. The average values for the tasks are 1.08, 1.42, 1.08, and 4.08, respectively. For the first three tasks, most participants were able to operate very close to the optimal strategies, i.e., without going back. For the last task~(a range chart), most participants more frequently used the Back button, trying to find the correct mappings between visual channels and data fields.


Overall, participants were satisfied with the usability of Mystique. P8 stated that ``\textit{it makes new visualizations with my dataset very easily by taking an example visualization found somewhere and adjusting the visualization. Also, it is very intuitive.}'' P8 also commented on the non-programming authoring process: ``\textit{It is a very interesting tool and useful for people who have no interest in programming. It might be useful for quick prototyping for visualization ideas.}'' P1 liked Mystique's wizard interface: ``\textit{Mystique gives a response after each step so I know whether I am on the right track, while in Python I cannot imagine what chart I am getting when writing codes there}.''

Participants also suggested several aspects for improvement. First, the guidance information in the Instruction Panel can be conveyed more clearly,
e.g., P8 mentioned ``\textit{it was hard to understand what x position exactly means in Task 4}'' and P6 commented ``\textit{the terms were too difficult to understand. It was hard to find what they meant exactly~(top side, bottom side, height, etc.).}'' Second, since customization is done after exporting the chart and not included in the study,
the participants wanted the reuse UI to support fine-tuning:
``\textit{one improvement could be to allow more flexibility; for instance, currently there is no option for selecting the color used in the chart}'' (P7).

\section{Discussion, Limitations, and Future Work}
\label{sec:discussion}


\noindent \textbf{Composite Designs.} The current deconstruction algorithm in Mystique cannot fully handle composite visualizations \cite{javed2012exploring} involving 
superimposition (e.g., \cref{fig:errors}), 
juxtaposition, overloading, and nesting.
Future work needs to extend the deconstruction algorithm or devise new methods to handle composite designs. For instance, given a design consisting of multiple views, we first need to dissect it into multiple visualizations. It remains to be seen how existing techniques on decomposing complex figures \cite{jiang_two-stage_2021,lee_detecting_2015,shi_layout-aware_2019} perform on real-world SVG visualizations.

\bpstart{Handling More Deconstruction Errors}
Mystique handles the potential errors and uncertainties in encoding detection by letting users choose the correct visual channel from a drop-down menu. As future work expands the scope to handle more complex charts like composite designs, it is expected that more errors will arise in the deconstruction process. How to support the user's understanding and provide ways to correct these errors remains an open problem. The challenge here is to minimize the requirement of the users' knowledge of abstract concepts and operations related to the GREC-component model. To do so, we need to have a thorough understanding of the error space, and then devise representation and interaction mechanisms to let users provide input in ways they can understand and perform.

\bpstart{\markup{Algorithmic Layouts}} \markup{Currently Mystique is unable to detect variants of the packing relationship. For instance, Mystique cannot tell apart a squarified treemap layout~\cite{bruls2000squarified} from a spatially ordered treemap layout~\cite{shneiderman2001ordered,wood2008spatially}. Even if such layout differences are known, the underlying library Mascot.js cannot reproduce the required layout yet. More investigations are necessary to accommodate those cases.}


\bpstart{\markup{Corpus Generalizability}} \markup{
We manually collected the corpus for evaluation to ensure diversity~\cite{chen2023state}, but the corpus size is small and may not be sufficient for further investigations of alternative deconstruction and reuse methods (e.g., neural network models). 
The corpus can be augmented in the following two aspects to support future research: (1) incorporating charts composed of other types of marks to enhance diversity and expressiveness and (2) for each chart type, increasing the number of charts evenly across different tools/sources.
}

\bpstart{Additional Features} Beside the research challenges outlined above, Mystique can benefit from a few feature enhancements. 
The user study revealed that people wanted to customize the design while performing the reuse steps~(e.g., categorical label ordering, the color set in the legend). Such functionality can be integrated into the reuse UI. 
Handling bespoke axis design is another potential future improvement. Mystique currently generates simple axes automatically based on detected axis information, and needs to support customizations such as label formatting, flipped axis, and dual axes (e.g., \cref{fig:teaser}(h)).
Finally, the visual style information of an online SVG chart sometimes is not embedded inside the SVG file, but stored in the web page or even a separate style sheet.
Capturing such visual styles is currently a manual process.
We expect that a future version of Mystique can support the automatic capturing of SVG charts and associated visual styles with simple interactions directly in the browser. 

\section{Conclusion}
Reusing existing charts with layouts determined by multiple factors has the potential to significantly transform the process and experience of visualization authoring, further lowering barriers to crafting bespoke charts. In this paper, we contribute Mystique, an interactive tool that brings us closer to realizing that potential. Mystique automates axis \& legend detection and chart deconstruction, and asks for minimal human input to fix detection errors and specify data mappings. We demonstrate that our mixed-initiative deconstruction approach can achieve above 96\% accuracy on charts with diverse layouts and designs. 
In addition, a chart reproduction study with 12 participants demonstrates that the guided process in Mystique's wizard interface is easy to follow and removes the need to learn a new language or framework. We outline research challenges and opportunities for future work, including broadening the scope to composite designs and enhancing Mystique with additional features.

\acknowledgments{%
	
Chen Chen and Zhicheng Liu were supported in part by NSF grant IIS-2239130, and Yunhai Wang was supported by NSFC (No. 62132017, 62141217) and Shandong Provincial Natural Science Foundation (No. ZQ2022JQ32).
}

\bibliographystyle{abbrv-doi-hyperref}

\bibliography{main}

\appendix 

\section{About Appendices}
Refer to \cref{sec:appendices_inst} for instructions regarding appendices.

\section{Troubleshooting}
\label{appendix:troubleshooting}

\subsection{ifpdf error}

If you receive compilation errors along the lines of \texttt{Package ifpdf Error: Name clash, \textbackslash ifpdf is already defined} then please add a new line \verb|\let\ifpdf\relax| right after the \verb|\documentclass[journal]{vgtc}| call.
Note that your error is due to packages you use that define \verb|\ifpdf| which is obsolete (the result is that \verb|\ifpdf| is defined twice); these packages should be changed to use \verb|ifpdf| package instead.

\subsection{\texttt{pdfendlink} error}

Occasionally (for some \LaTeX\ distributions) this hyper-linked bib\TeX\ style may lead to \textbf{compilation errors} (\texttt{pdfendlink ended up in different nesting level ...}) if a reference entry is broken across two pages (due to a bug in \verb|hyperref|).
In this case, make sure you have the latest version of the \verb|hyperref| package (i.e.\ update your \LaTeX\ installation/packages) or, alternatively, revert back to \verb|\bibliographystyle{abbrv-doi}| (at the expense of removing hyperlinks from the bibliography) and try \verb|\bibliographystyle{abbrv-doi-hyperref}| again after some more editing.

\end{document}